\documentclass[%
 aip,
 amsmath,amssymb,
 reprint,%
]{revtex4-1}
\usepackage{xcolor} 

\usepackage{graphicx}
\usepackage{bm}

\usepackage[utf8]{inputenc}
\usepackage[T1]{fontenc}
\usepackage{mathptmx}
\usepackage{etoolbox}
\usepackage{upgreek}
\usepackage{float}

\makeatletter
\def\@email#1#2{%
 \endgroup
 \patchcmd{\titleblock@produce}
  {\frontmatter@RRAPformat}
  {\frontmatter@RRAPformat{\produce@RRAP{*#1\href{mailto:#2}{#2}}}\frontmatter@RRAPformat}
  {}{}
}%
\makeatother
\begin{document}

\preprint{AIP/123-QED}

\title[]{Fast measurement of group index variation with optimum precision using Hong-Ou-Mandel interferometry}

\author{Sandeep Singh}
\email{realacumen@gmail.com}
\affiliation{Photonic Sciences Laboratory, Physical Research Laboratory, Navarangpura, Ahmedabad, 380009 Gujarat, India}
\affiliation{Indian Institute of Technology-Gandhinagar, Ahmedabad 382424, Gujarat, India}

\author{Vimlesh Kumar}%
\affiliation{Photonic Sciences Laboratory, Physical Research Laboratory, Navarangpura, Ahmedabad, 380009 Gujarat, India
}%

\author{G. K. Samanta}
\affiliation{Photonic Sciences Laboratory, Physical Research Laboratory, Navarangpura, Ahmedabad, 380009 Gujarat, India}%

\date{\today}

\begin{abstract}
Hong-Ou-Mandel (HOM) interferometry has emerged as a valuable means for quantum sensing applications, particularly in measuring physical parameters that influence the relative optical delay between photon pairs. Unlike classical techniques, HOM-based quantum sensors offer higher resolution due to the intrinsic dispersion cancellation property of correlated photon pairs. Due to the use of single photons, the HOM-based quantum sensors typically involve a large integration time to acquire the signal and subsequent post-processing for high-resolution measurements, restricting their use for real-time operation. Based on our understanding of the relationship between measurement resolution and the gain medium length that produces photon pairs, we report here on the development of an HOM-based quantum sensor for high-precision group index measurement. Using 1-mm long periodically-poled KTP (PPKTP) crystal for photon-pair generation, we have measured group index with a precision of $\sim 6.75\times 10^{-6}$ per centimeter of sample length at an integration time of 100 ms, surpasses the previous reports by 400$\%$. Typically, the measurement range reduces with the increase of the resolution. However, using a novel scheme compensating photon delay due to group index changes stepwise with an optical delay stage, we have measured the group index variation of PPKTP crystal over a range of 3.5 $\times 10^{-3}$ for a temperature change from 25°C to 200°C, corresponding to an optical delay adjustment of approximately 200 $\upmu$m while maintaining the same precision ($\sim$6.75$\times 10^{-6}$ per centimeter of sample length). The current results establish the usefulness of HOM-interferometer-based quantum sensors for fast, precise, and long-range measurements in various applications, including quantum optical coherence tomography.

\end{abstract}

\maketitle

\section{Introduction}
Quantum optical coherence tomography (QOCT) represents a non-invasive imaging modality primarily employed to generate three-dimensional cross-sectional visualizations of transparent or reflective materials, specifically focusing on the microstructural surface morphology of biological tissues\cite{nasr2003, povazay2002}. In QOCT, unlike conventional optical coherence tomography with classical light, the utilization of entangled photon pairs allows probing of the sample surface with an axial resolution enhanced by a factor of 2 even for the same spectral bandwidth source in both scenarios, showcasing the quantum advantage in precision imaging. On the other hand, the immunity against the dispersion effects owing to the frequency-entangled photon pairs enables QOCT to access higher resolution of surface morphology at sub-micron scales\cite{nasr2003} through the use of a broader spectrum of photons compared to classical counterparts.
The conventional QOCT relies on the Hong-Ou-Mandel (HOM) interference, where two indistinguishable photons coalesce at a balanced beam splitter, with one photon traversing a reference delay line and the other probing the sample surface, encoding its features in the relative temporal delay between the photon pairs \cite{PhysRevLett.59.2044, kashi2023, alsing2022}. Since changes in temporal delay lead to variations in coincidence counts and the surface morphology of the specimen influences this delay, one can map the surface morphology by recording the coincidence counts in the transverse plane. However, the axial resolution of such measurements is fundamentally limited by the coherence length (time) of the indistinguishable photons, thus tied to the width of the HOM interference dip. The shape of the HOM dip, essentially the Fourier transform of the spectral profile of the photon pairs, indicates that a broader spectral bandwidth of the photon pairs results in a narrower HOM dip. This relationship underscores the critical role that the spectral characteristics of the photon pairs play in determining the axial resolution achievable in QOCT measurements. For example, in the case of a stratified specimen, the probe photon acquires data pertaining to the optical thickness across the distinct layers. As a result, the precise determination of the thickness profile of individual layers demands a meticulous understanding of their corresponding group indices. Even in quantum optical coherence microscopy (QOCM), achieving an accurate three-dimensional reconstruction of the specimen necessitates a precise determination of the group index variations within the transverse plane of the sample. Unfortunately, these group index variations, in general, are exceedingly small, and conventional methodologies for group index measurement lack the necessary precision to measure those subtle changes. 
Nevertheless, while HOM interference has been applied for measuring group index through the assessment of optical delay \cite{lyons2018attosecond} introduced by the sample and has enabled high-precision measurements in quantum microscopy \cite{ndagano2022}, QOCT \cite{abouraddy2002}, and vibration sensing \cite{singh2023AQT} without adequately addressing the crucial requirement to improve the precision of HOM interferometry. Typically, achieving high precision requires stringent conditions, such as long interaction lengths of the specimen and subsequent post-processing of large data accumulated iteratively \cite{reisner2022}. As a result, such approaches suffer from the common drawback of long measurement time and limited kinds of samples. For example, the measurement of group index variation of the nonlinear crystals due to the temperature-dependent dispersion properties is of paramount interest for designing various experiments, including the optimization of phase-matching for high nonlinear conversion efficiency and optimum phase-compensation of the entangled state generated by the nonlinear crystal. However, the precision measurement using the existing HOM interferometer-based group index measurement techniques demands long interaction lengths. 
Unfortunately, in many cases, nonlinear crystals exhibit complicated growth techniques \cite{zhou2023}, inherent fragility, and high cost \cite{dhanaraj2010}, posing challenges in procuring samples with extended interaction lengths. Therefore, it is imperative to enhance the resolution of the HOM interferometer. In our recent study, we demonstrated that the resolution of the HOM interferometer can be improved by increasing the spectral bandwidth of photon pairs, achieved through the appropriate selection of crystal length for generating these photon pairs \cite{singh2023AQT}. 
In this study, we build upon the rigorous characterization of the HOM interferometer and the strategies to enhance its measurement precision, as reported in ref \cite{singh2023AQT}. Employing a 1 mm long periodically-poled potassium titanyl phosphate (PPKTP) crystal\cite{Singh23b} to generate photon pairs with a spectral bandwidth of 162.05 $\pm$ 1.12 nm, 
we developed a HOM interferometer-based quantum sensor to measure the temperature-dependent group index variation of another PPKTP crystal with a resolution of approximately $2.25 \times 10^{-6}$ over a measurement range of approximately $3.5 \times 10^{-3}$ by varying crystal temperature across  $175^\circ$C. 
We achieved an unprecedented improvement in the group-index measurement resolution, a 400\% improvement as compared to previous reports. This 1 mm long PPKTP crystal provides a high photon-pair generation rate with sufficiently broad spectral bandwidth, enabling optimum precision in group index measurements while maintaining fast acquisition speeds, with exposure time as short as 50 ms. Additionally, access to such a high resolution does not require any post-processing, enabling the possibility of real-time quantum sensing of any physical parameter influencing the optical delay between the photon pairs. Further improvement in the measurement resolution can be achieved by increasing the spectral bandwidth of the photon pairs by further reducing the length of the gain medium (PPKTP crystal), however, at the cost of lower generation of photon pairs rate and increased measurement time.    

The group index ($n_{g}$) of a dispersive material is related to its refractive index ($n$) as follows \cite{miriampally2016},
\begin{equation}
\begin{split}
    n_{g} & = \frac{c}{v_{g}}\\
          & = n - \lambda\frac{dn}{d\lambda}|_{\lambda = \lambda_{\circ}}\\
\end{split}
\label{eq1}
\end{equation}
where, c, and $v_{g}$ are the velocity of light in vacuum and the group velocity of light in the material, respectively. The 
$\lambda_{\circ}$ is the central wavelength of the photon wave packet traveling through the dispersive material. It is evident 
from Equation \ref{eq1} that in the normal dispersion regime, the group index of a material, $n_{g}$, is always higher than 
its refractive index, $n$. The optical delay, $\Delta x$, associated with the group index of the material for an effective interaction length of $L$ can be expressed as,
\begin{equation}
    {\Delta x = n_{g}L}\\
\label{eq2}
\end{equation}
Equation \ref{eq2} makes it clear that the precision of measuring the group index, $n_{g}$, of a specimen is contingent upon the precision in measuring the optical delay introduced by it and its physical length, $L$. Hence, accessing the ultimate precision, which is the central goal of this report, in optical delay measurement becomes imperative to accurately measure the group delay introduced by the material under test for quantum imaging and microscopy.

\section{Experimental setup}

\begin{figure*}[ht]
\centering
\includegraphics[width=0.9\linewidth]{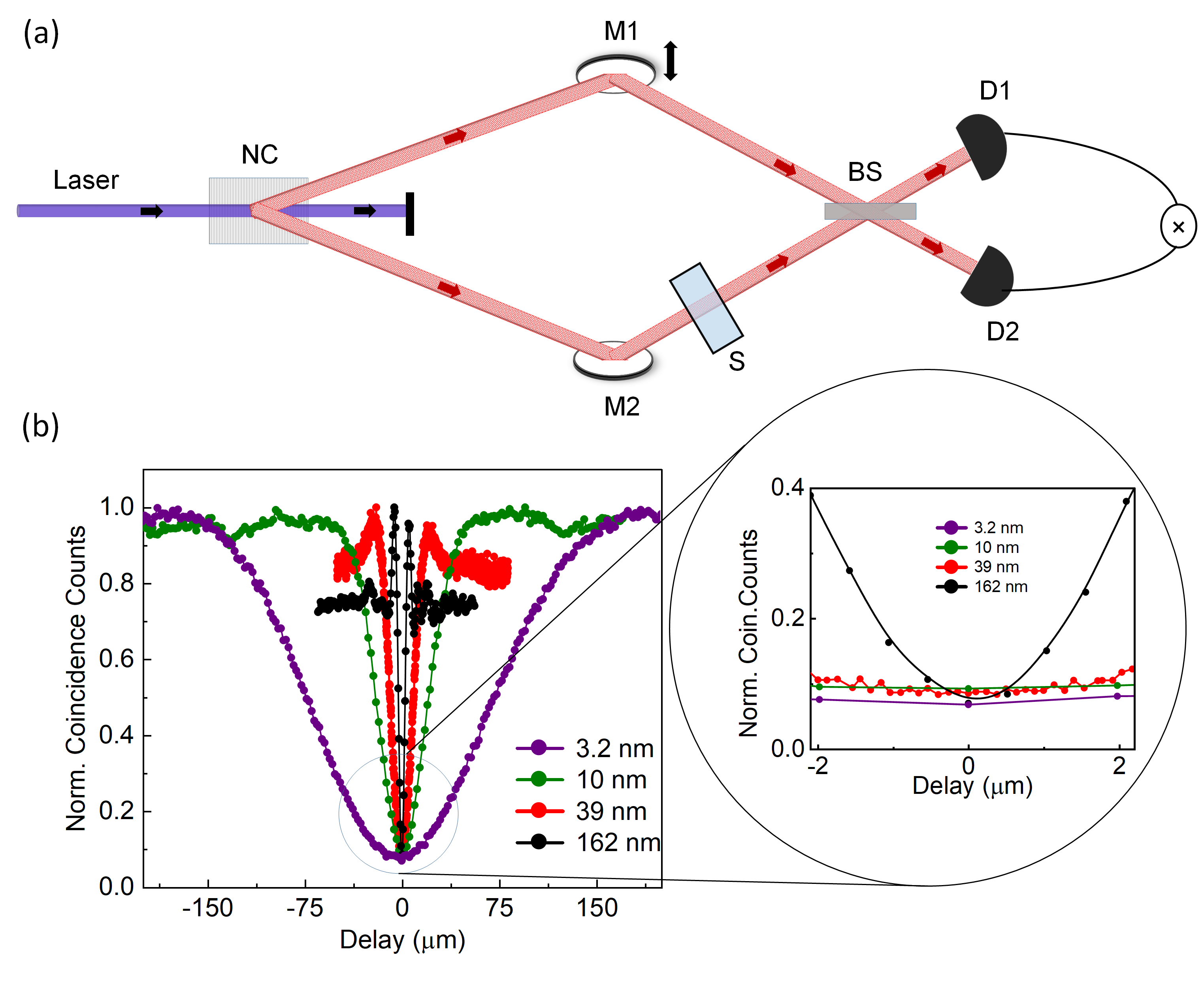}
\caption{a) Schematic of the experimental setup. \textbf{NC}: PPKTP crystal of grating period = 3.425 $\upmu$m, \textbf{S}: Specimen, \textbf{M1-2}: dielectric mirrors, \textbf{BS}: 50:50 beam splitter, \textbf{D1-2}: Single photon counting modules. b) HOM interference profiles for photon pairs of different spectral bandwidths. (Inset) Magnified image of the HOM dips. }
\label{Figure 1}
\end{figure*}

The schematic of Fig. \ref{Figure 1}(a) shows the concept of quantum sensing using HOM interferometry, while the detailed experimental scheme is the same as reported in Ref. \cite{singh2023AQT}. A single-frequency, linearly polarized (vertical polarization), continuous-wave (CW), fiber-coupled diode laser providing 20 mW of output power at 405.4 nm is used as the pump laser in the experiment. 
The pump laser is focused at the center of the nonlinear crystal for the optimum rate of photon-pair generation. Two periodically poled KTiOPO$_4$ (PPKTP) crystals\cite{VKumar22, Singh23b} of the same aperture (1 $\times$ 2 mm$^2$) but of different interaction lengths, $L$ = 1, and 20 are used as the nonlinear crystals (NC) for the experiment. Both the nonlinear crystals have a single grating period of $\Lambda$ = 3.425 $\upmu$m designed for the degenerate, type-0 ($e $→$ e + e$) phase-matched parametric down-conversion (PDC) of 405 nm to 810 nm. To set out quasi-phase-matching, the PPKTP crystals are kept in an oven having dynamic temperature control over room temperature to 200$^{\circ}$C with temperature stability of $\pm$0.1$^{\circ}$C. After being separated from the pump, the non-collinear, down-converted photon pairs are collimated and guided in two different paths to interfere with the balanced (50:50) beam splitter (BS). The relative optical delay between the photon pairs is adjusted using the optical delay line in the path of one of the photons. On the other hand, the sample, S, under study is placed in the path of another photon. The photons at the output ports of the BS are collected using single-mode fiber connected to the single-photon counting modules, D1 and D2 (AQRH-14-FC, Excelitas). Further, the time-to-digital converter, TDC (ID-800), is used to measure the single and coincidence counts. All optical components used in the setup are selected for optimal performance at both the pump and PDC wavelengths. All the data is recorded at a coincidence window of 1.62 ns, integration time of 20 ms, 50 ms, and 100 ms, and a typical pump power of 0.25 mW. The 1.62 ns coincidence window was selected to balance the effects of electronic jitter, which reduces coincidence counts at narrower windows, and the detection of uncorrelated photon pairs at wider windows, thus minimizing both the constraints to ensure accurate measurements. \\

\section{Results and Discussions}
In this study, we specifically opted for degenerate, type-0 phase-matching configuration, which offers its highest photon-pair generation rate in collinear configuration. However, it creates signal and idler photons with the same wavelength and polarization, making it difficult to separate them. The practical approach to separating the correlated photon pairs using a 50:50 beam splitter reduces coincidence counts by 50$\%$\cite{yepiz2022}, effectively negating the benefit of the high photon flux offered by this particular phase-matching configuration. On the other hand, type-0, non-collinear phase-matching enables easy separation of the signal and idler photons as they are distributed at diametrically opposite points over an annular ring but at the cost of a lower collection rate. This collection rate can be improved by reducing the annular ring size of the photon pairs \cite{Jabir:17}.
\\
We first studied the performance of the HOM interferometer for indistinguishable photons of four different spectral bandwidths. To verify the optimum performance of the collection optics not affecting the spectral bandwidth of the detected photons\cite{Varga2022}, we experimentally measured the spectral bandwidth of the photon pairs using a spectrometer (HR-4000, Ocean Optics) to be 30 nm and 37 nm for 30 and 20 mm long crystals, respectively. Such measurement was possible due to the high photon pairs rate arising from the high figure-of-merit of the PPKTP crystal in degenerate, type-0, phase-matching conditions. The experimentally measured spectral bandwidth is found to be in close agreement with the spectral bandwidth estimated from the HOM interferometry, confirming the reliability of the collection systems.
It is evident from Ref. \cite{singh2023AQT} that the photon pairs generated from PPKTP crystal of length, L = 1 mm, and 20 mm, have spectral bandwidths of 162.05 $\pm$ 1.12 nm, and 39.08 $\pm$ 0.19 nm, respectively. To further reduce the spectral bandwidth, we have used interference filters with Gaussian transmission bandwidths of 3.2 nm and 10 nm centered at 810 nm. Using the photon pairs with four different spectral bandwidths, 162 nm, 39 nm, 10 nm, and 3.2 nm, all centered at 810 nm, we measured the HOM interference by recording the coincidence counts with the relative optical delay. The results are shown in Fig. \ref{Figure 1}(b). The FWHM width and HOM visibility of the HOM dips are found to be (4.06 $\pm$ 0.01 $\upmu$m, $\sim$92.8\%), (16.90 $\pm$ 0.03 $\upmu$m, $\sim$91.7\%), (42.40 $\pm$ 0.06 $\upmu$m, $\sim$90.7\%) and (144.51 $\pm$ 0.08 $\upmu$m, $\sim$93.2\%) for spectral bandwidth of 162.05 $\pm$ 1.12 nm, 39.08 $\pm$ 0.19 nm, 10 $\pm$ 0.01 nm, and 3.2 $\pm$ 0.01 nm, respectively. The HOM visibility was calculated using the formula, $V_{HOM} = \frac{C_{max} - C_{min}}{C_{max}}$, where $C_{max}$ and $C_{min}$ are the maximum and the minimum coincidence counts on the HOM dip \cite{bouchard2020}. We observe an inverted Gaussian function approximates the HOM curve (purple line and dots) for the photon pairs with the spectral bandwidth of 3.2 and 10 nm, demonstrating the Gaussian transmission profile of the narrow bandpass filter used for spectral filtering. For broadband photon pairs of spectral bandwidths 39 nm and 162 nm were obtained by removing the pump using high-pass filters, the HOM curves show a profile closely approximated by an inverted sinc function. The oscillatory behavior of coincidence counts lying on either side of the interference dip results from spatial beating due to interference of non-overlapping spectral frequencies within the broadband two-photon wave packet, while damping occurs due to the lower probability of higher spectral frequencies \cite{biphotonbeat2019, ou1988}. To gain further insights, we have enlarged the minimum coincidence section of the HOM curves, as depicted in the inset of Fig. \ref{Figure 1}(b). Notably, within the optical delay range of $\pm$2 $\upmu$m, the minimum coincidence counts for photon pairs with spectral bandwidths of 3.2 nm, 10 nm, and 39 nm coincide, leading to higher inaccuracies in the measurement of physical parameters (e.g., the refractive index of a sample) utilizing the shift of HOM dip. On the other hand, the slope (change in coincidence with optical delay) of the HOM curve for photon pairs with broad spectrum (black dots) significantly enhances measurement accuracy. Ideally, the slope of the HOM curve is constant for the photon pairs of fixed spectral bandwidth. However, in practice, HOM visibility, which is majorly affected by imperfect spatial overlap of the pair photons, the beam splitter, and multi-photon events due to high parametric gain, can reduce the slope of the HOM curve. Take, for instance, the difference in HOM visibility between the 1 mm and 20 mm PPKTP crystals can be understood not only in terms of changes in the spectral bandwidth of the photon pairs generated by each crystal but also due to the altered spatial overlap of photon pairs on the beam splitter while switching between the crystals used for pair photon generation. This causes slight changes in the optical alignment of pair photons on the beam splitter.
Although precise alignment and spectral filtering can improve visibility, spectral filtering can also reduce the width of the HOM dip and photon flux. Therefore, it is essential to optimize all these key parameters to increase the slope of the HOM curve in achieving the highest measurement precision.\\
To increase the slope, we have achieved the narrowest HOM curve (width of 4.06 $\pm$ 0.01 $\upmu$m) and high visibility ($\sim$92.8\%) using a PPKTP crystal of 1 mm interaction length. Arguably, crystal lengths shorter than 1 mm could increase spectral bandwidth and measurement resolution at the cost of a lower photon pair generation rate and measurement speed. Although increasing the pump power can boost the generation rate, it also leads to a rise in multi-photon events, which reduces HOM visibility and, consequently, the measurement resolution. In our experiments, we observed that increasing the crystal length from 1 mm to 30 mm resulted in a decrease in the slope of the HOM curve, from 450 $\upmu$m$^{-1}$ to 260 $\upmu$m$^{-1}$, at a pump power of 0.25 mW and an integration time of 20 ms \cite{singh2023AQT}. Therefore, the use of a 1-mm long PPKTP crystal ensures an optimal balance to maintain high resolution and fast measurement. Due to the unavailability of crystals of length shorter than 1 mm, we could not observe the effect of smaller crystal lengths on the spectral bandwidth and HOM curve experimentally.

\begin{figure*}[ht]
\centering
\includegraphics[width=\linewidth]{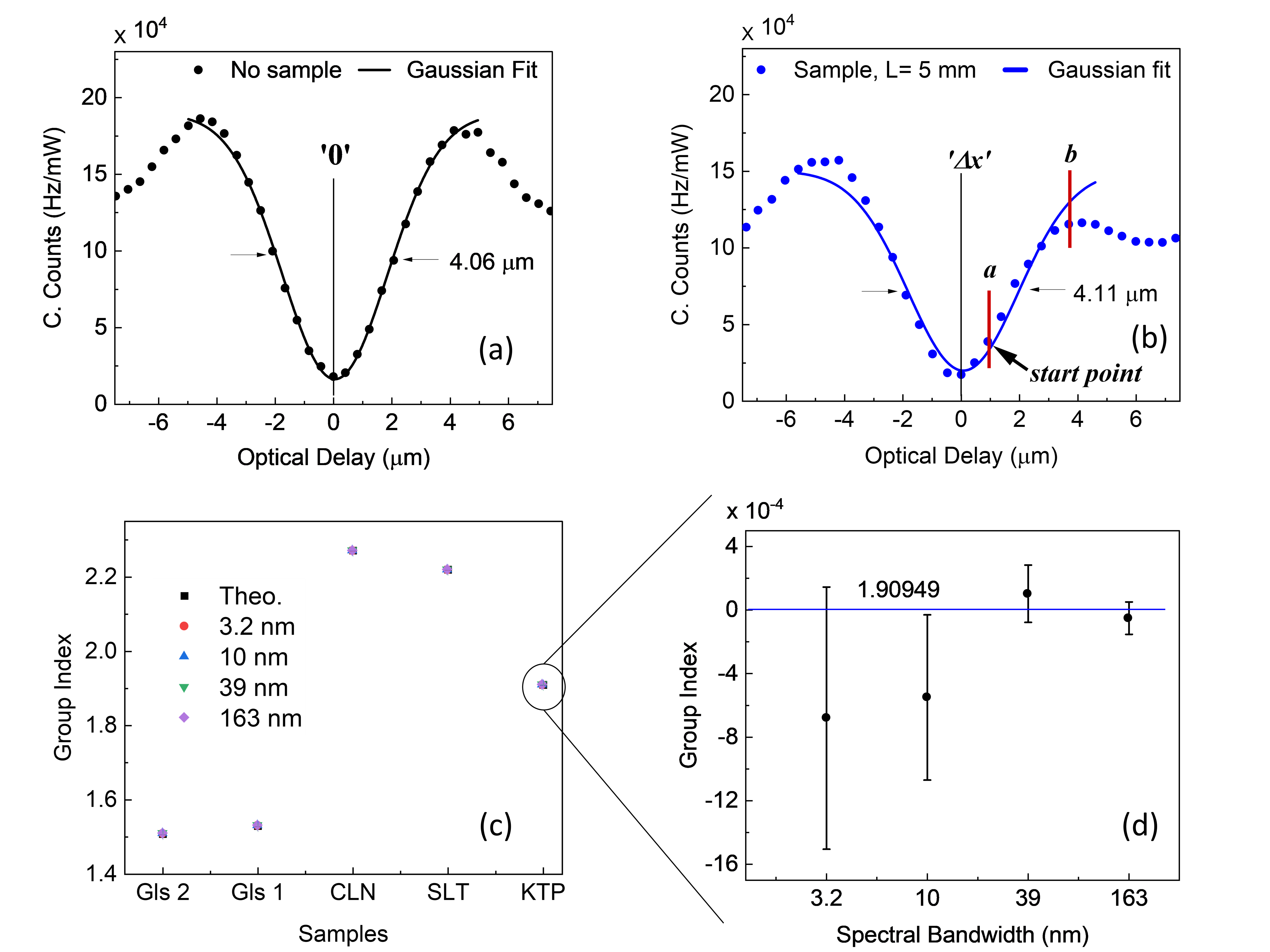}\
\caption{Variation of coincidence counts with optical delay of HOM interferometer for photon pairs of the spectral bandwidth of 162 nm in a) absence and b) presence of the sample of physical length, $L$ = 5 mm. Solid lines represent the theoretical fit to the experimental data denoted by solid dots c) Group index of five samples measured using HOM interferometer for photon pairs of different spectral bandwidths. d) Variation of group index of KTP crystal of constant length and temperature as a function of spectral bandwidth of the photon pairs.}
\label{Figure 2}
\end{figure*}
Furthermore, to demonstrate the necessity of a narrow HOM curve for high-resolution HOM-based quantum sensors, we employed HOM interferometry to measure the optical delay between photon pairs induced by a dispersive material with varying group indices. Utilizing the measured group index and the corresponding dispersion relations, one can also deduce the refractive index of the material \cite{miriampally2016}. For group index measurement, we have used five different samples (S), including the commonly used periodically poled nonlinear crystals, KTP, Mgo-doped stoichiometric grown lithium tantalate (SLT), and MgO-doped congruent grown lithium niobate (CLN) of lengths, 5.07, 5.18, and 10.03 mm, respectively and two glass slabs, named as Gls1 (Schott glass) and Gls2 (BK7) of lengths, 3.10, and 1.15 mm, respectively. 
As expected, the insertion of the sample (S) in one of the arms of the HOM interferometer shifts the HOM curve, So we again retrieved the HOM curve using the delay stage employed in the other arm of the HOM interferometer. Subsequently, using the overall shift ($\Delta x$) of the interference minima due to the presence of the samples for all four spectral bandwidths of the photon pairs, we calculated the group index using Eq. \ref{eq2}. The results are shown in Fig. \ref{Figure 2}. As evident from Fig. \ref{Figure 2}(a), the photon pairs generated by the  PPKTP crystal of length 1 mm have a symmetric HOM curve, in the absence of any sample, of FWHM width (obtained using Gaussian curve fit of the experimental data, shown by solid black curve) of 4.06 $\pm$ 0.012 $\upmu$m. The minimum coincidence point of the HOM curve is used as the reference point and marked as "0". We record the linear stage position at this point as 21.554 mm. However, the insertion of the sample (PPKTP crystal of length, L = 5 mm) shifts the HOM curve, as shown in Fig. \ref{Figure 2}(b). The HOM curve has an asymmetry, but the FWHM width of 4.11 $\pm$ 0.03 $\upmu$m (obtained using Gaussian curve fit of the experimental data, shown by solid blue curve) is almost the same as the FWHM width of the HOM curve in the absence of the sample. Such observation confirms the dispersion cancellation feature of the HOM interferometer \cite{Okano:15} despite the broad spectral bandwidth (162 nm) of the pair photons. The minimum coincidence point of the HOM curve in the presence of a 5 mm long PPKTP sample is shifted to a new position marked as $\Delta x$ with a linear stage reading of 26.164 mm such that $\Delta x$ = ($n_g$-1)L = 4.61 mm. Using this optical delay; we can measure the group index of the sample. We repeated the same measurement procedure for all the available samples of different lengths and spectral bandwidths of the photon pairs and measured their group indices. The results are shown in Fig. \ref{Figure 2}(c). It is observed that the group indices of all the five samples measured (dots) using photon pairs of different spectral bandwidths match with their respective group indices calculated using the respective Sellmeier equation (black dot). Such an observation establishes the capability and reliability of the HOM interferometer as a high-precision refractometer to measure the group index of any dispersive materials. In this measurement, all the samples (S) were kept at room temperature.

To scrutinize the accuracy of the measured group indices under the influence of various parameters contributing to inaccuracies in the measurement of ($\Delta x$), especially while using different spectral bandwidths of photon pairs, we have amplified the scale of the recorded group index data for the PPKTP crystal (as an example). The results are shown in Fig. \ref{Figure 2}(d). As evident from Fig. \ref{Figure 2}(d), the group index (dots) of the PPKTP crystal measured using different spectral bandwidths of the photon pairs lies close to the theoretical value (blue line) calculated using the dispersion relations \cite{kato2002} of the PPKTP crystal. However, careful observation reveals that the measurement precision increases from $\pm 8.3 \times 10^{-4}$ to $\pm 1.0 \times 10^{-4}$, indicating an 8-fold enhancement when using HOM interference with photon pairs having a broad spectral bandwidth of 162 nm, compared to 3.2 nm. 
The observed enhanced precision in group index measurement is due to the improved accuracy in detecting shifts in the HOM curve with a narrow dip width, a result of using photon pairs with a broad spectral bandwidth. Additionally, the group velocity dispersion of the sample in the presence of different spectral bandwidths of photon pairs showed almost no change in measurement resolution, further confirming the dispersion cancellation characteristics of the HOM interferometer, consistent with the observation of two-photon interferometry-based QOCT using ultra-broadband photon pairs \cite{Okano:15}. From these results, it is evident that employing the shift of the entire HOM curve as a pointer enables the measurement of the group index of any dispersive medium with an accuracy in the range of $\pm 1\times10^{-4}$ when using photon pairs with a broad spectral bandwidth of 162.05 $\pm$ 1.12 nm. Consistent performance is evident in group index measurements across the remaining materials. For a better understanding, we have tabulated all the group index measurement parameters in Table \ref{table1}. The spectral bandwidth of the photon pairs used in this measurement is 162.05 $\pm$ 1.12 nm. As evident from Table \ref{table1}, the measurement accuracy of the group index of all the dispersive mediums with different lengths is expectedly in the range of $\pm 1\times10^{-4}$, similar to the order of precision achieved by conventional OCT techniques\cite{Singh:02, Yablon:13}. Like conventional OCT techniques, here, the overall precision of the group index depends on the measurement precision of the sample length also, as the phase difference or optical delay is the product of both the group index and length of the sample (see Eq. \ref{eq2}). Therefore, access to the ultimate precision of the group index measurement demands high precision of sample length through sophisticated techniques, e.g., electron microscopy\cite{villarrubia2015} over vernier calipers. Despite the high precision in measuring sample length, the intrinsic limitations of conventional OCT techniques, caused by normal dispersion, restrict their effectiveness for high-precision group index measurements in longer samples due to degraded temporal coherence and, thus, the interference fringes. Contrary to this, HOM interferometry utilizes the frequency correlation of entangled photon pairs to counteract the effects of group velocity dispersion (GVD) due to its inherent dispersion cancellation property \cite{steinberg1992dispersion}. As a result, the resolution in QOCT, represented by the width of the HOM dip, remains unaffected by the GVD. This makes the HOM interferometer-based current technique competent enough for the real-time measurement of subtle changes in group indices and useful for long sample lengths without requiring any post-processing of the acquired data \cite{reisner2022}.

However, even this measurement precision reported in Table \ref{table1} is likely insufficient for detecting subtle variations in group indices, In particular, the temperature-dependent group index (thermo-optic dispersion) of different nonlinear crystals particularly relevant for quasi-phase-matched nonlinear parametric processes. Commonly, the quasi-phase-matched crystals available in the market have a weak temperature-dependent refractive index variation in the range of $10^{-4}$ to $10^{-5}$ $^\circ$C$^{-1}$. Therefore, we need the precision at least one order of magnitude higher, i.e., $\sim10^{-6}$. However, Recent advancements have demonstrated the ability to measure a static optical delay of $\sim$ 60 nm using HOM interferometry with photon pairs of broad spectral bandwidth \cite{singh2023AQT}, underscoring the potential of this method to fulfill the need for necessary resolution in temperature-dependent group index measurements. As the optical delay is defined as the product refractive index, $n$, and physical length, $L$, of a medium, i.e., $n_{g}\times L$, it is feasible to measure changes in the group index smaller than $\sim$10$^{-6}$ by carefully selecting the physical length of the medium. 

\begin{table*}[ht]
\caption{\bf Group index ($n_{g}$) of different materials at room temperature.}
\begin{ruledtabular}
\begin{tabular}{cccccc}
Sample & Length  & Theory & Measured  & Accuracy & Sellmeier \\
 &  (mm) &  $n_{g}$ &  $n_{g}$ &   & equation \\
\hline
PPKTP crystal & 5.07 & 1.90949 & 1.90944 & 1.02 $\times$ $10^{-4}$ & \cite{kato2002}\\
PPSLT crystal & 5.18 & 2.21997 & 2.21994 & 8.97 $\times$ $10^{-5}$ & \cite{manjooran2012}\\
MCLN crystal & 10.03  & 2.27129 & 2.27132 & 8.04 $\times$ $10^{-5}$ & \cite{manjooran2012}\\
Schott Glass (Gls1) & 1.15 & 1.52927 & 1.52924 & 1.05 $\times$ $10^{-4}$ & \cite{rii}\\
Glass slab (Gls2) & 3.10 & 1.50778 & 1.50773 & 8.97 $\times$ $10^{-5}$ & \cite{GAN1995}\\
\end{tabular}
\end{ruledtabular}
\label{table1}
\end{table*}
To validate the feasibility of measuring the temperature-dependent group index (thermo-optic dispersion) of periodically poled crystals, we selected a PPKTP crystal. This choice is advantageous for two reasons: firstly, experimental observations can be corroborated using standard data from available Sellmeier equations \cite{kato2002}, and secondly, the PPKTP crystal exhibits approximately 15 times less susceptibility to temperature fluctuations compared to other popular periodically poled crystals such as lithium niobate and lithium tantalate \cite{manjooran2012, lin2017}. This makes PPKTP an optimal selection for demonstrating the efficacy of this configuration in discerning subtle variations in group indices.

However, before conducting the experimental observations, we analytically calculated the variations in the group index of the PPKTP crystal as a function of crystal temperature. Let us assume that $x(T_{\circ})$ is the optical delay between the photon pairs due to the PPKTP crystal of length $L$ and group index $n_g$ at the initial temperature $T_{\circ}$. The $x(T_{\circ})$ can be written as:
\begin{equation}
x(T_{\circ}) = n_{g}(T_{\circ})\times L(T_{\circ})
\label{eq3}
\end{equation}
Now, the change in crystal temperature by $\Delta T$ leads to the change in its group index and effective interaction length due to thermo-optic dispersion and thermal expansion, respectively. Therefore, the temperature-dependent optical delay, $x(T_{\circ}+\Delta T)$, can be written as,
\begin{equation}
    x(T_{\circ}+\Delta T) = (n_{g}(T_{\circ}) + \Delta n_{g}(\Delta T))\times(L(T_{\circ})+\Delta L(\Delta T))
\label{eq4}    
\end{equation}
where, $\Delta n_{g}(\Delta T)$ and $\Delta L(\Delta T)$ denote the change in group index and the effective interaction length of the crystal, respectively, due to the change in crystal temperature of $\Delta T$. The net change in the relative optical delay corresponding to the change in crystal temperature can be written as,
\begin{equation}
    \Delta x(\Delta T) = x(T_{\circ}+\Delta T) - x(T_{\circ})
\label{eq5}    
\end{equation}
Using Eq. \ref{eq3} and Eq. \ref{eq4} in Eq. \ref{eq5} and neglecting the term $\Delta n_{g}(\Delta T)\Delta L(\Delta T)$ due to its small magnitude ($\sim  10^{-10}$) as compared to the measurement accuracy, we can represent the temperature-dependent group index change as,
\begin{equation}
     \Delta n_{g}(\Delta T) \simeq \frac{\Delta x (\Delta T) - n_{g}\Delta L(\Delta T)}{L}
\label{eq6}     
\end{equation}
The second term in the numerator of Eq. \ref{eq6} accounts for the change in relative optical delay between the signal and idler of the photon pairs due to the thermal expansion of the sample PPKTP crystal. Since the refractive index and crystal lengths vary simultaneously with the temperature, we can measure one of the parameters at a time for a given value of the other parameter. Therefore, we used the available thermal expansion data of PPKTP crystal from the literature using Ref. \cite{emanueli2003} and measured the temperature-dependent group index variation of the PPKTP crystal. As evident from Eq. \ref{eq6}, for a given crystal length, L, by measuring the change in optical delay, $\Delta x(\Delta T)$, one can easily find the temperature-dependent group index of the crystal. In doing so, we have reconstructed the HOM curve in the presence of the sample (PPKTP crystal of measured length, L = 30.12 mm) at a fixed temperature of 26$^{\circ}$C. Given the intrinsic dispersion cancellation property as observed in Fig. \ref{Figure 2} and Ref. \cite{Okano:15}, we obtained the HOM curve of FWHM width same as the HOM curve presented in Fig. \ref{Figure 2}(b)), despite having such a long sample. 
\begin{figure}[ht]
    \centering
    \includegraphics[width=\linewidth]{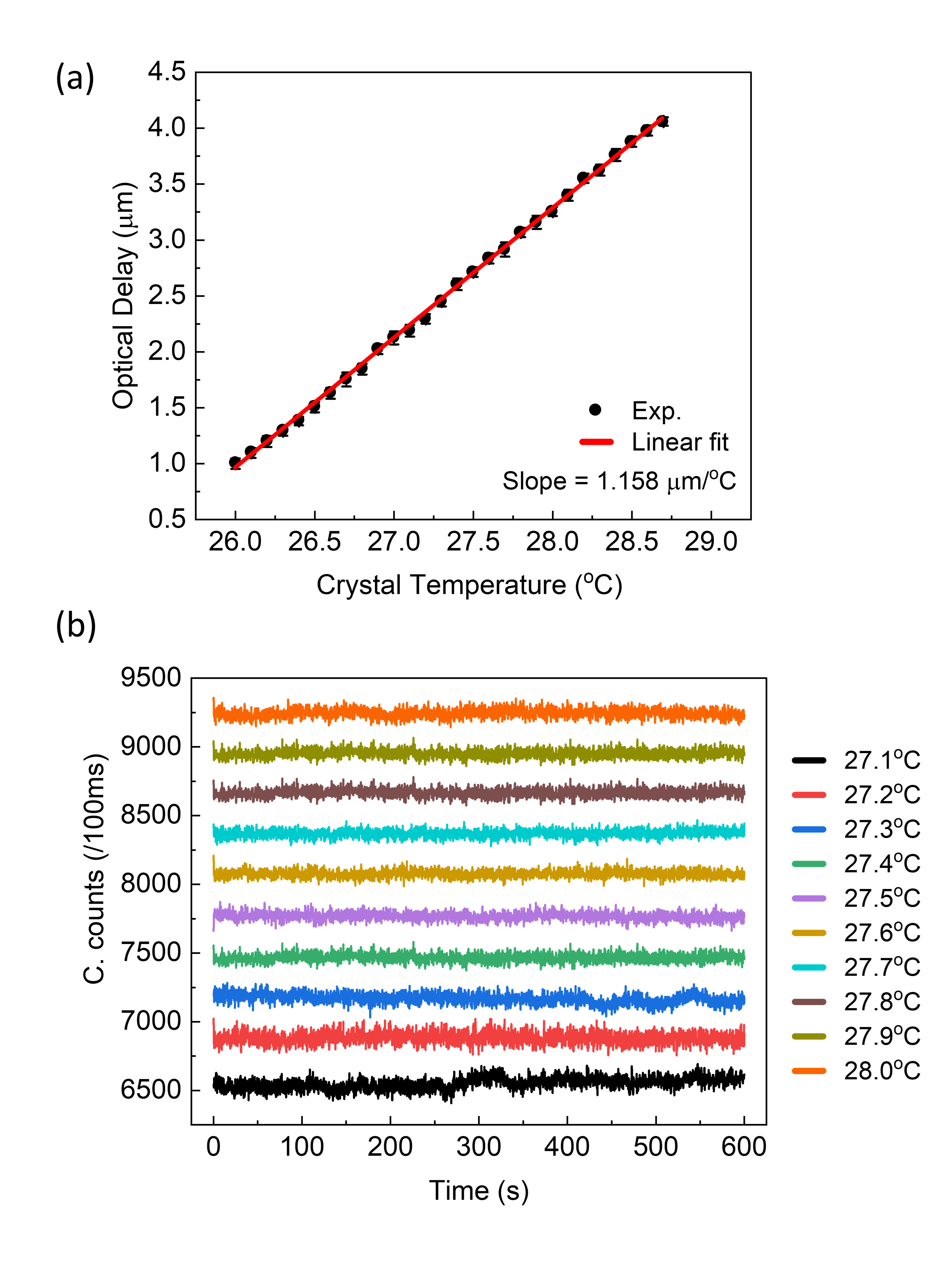}
    \caption{a) Variation of optical delay as a function of crystal temperature measured using the delay-dependent coincidence counts variation of HOM curve (see Fig. \ref{Figure 2}). The straight line (red) is the linear fit to the experimental results. b) Temporal stability of the coincidence counts for a period of 10 minutes at each increment (1$^{\circ}$C) in the sample crystal temperature.}
    \label{Figure 3}
\end{figure}

For calibration, we measured the optical delay due to the change in the group index of the crystal as a function of its temperature. First, we adjusted the delay line so that the initial point (named the start point, see Fig. \ref{Figure 2}(b)) corresponds to an optical delay of +1 $\upmu$m. The crystal temperature corresponding to the start point is $T_{\circ} = 26^\circ$C. We have also identified the measurement range by two red lines labeled as a (coincide with start point) and b (final point). This selection is purposefully made to access most of the linear range of the HOM curve, thereby enhancing the measurement range. Subsequently, we changed the temperature of the sample (PPKTP crystal) in minimum steps of 0.1$^\circ$C, as allowed by the oven. Since these small increments are close to the stability range of the oven, for each temperature rise of 0.1$^\circ$C, we allowed the oven and crystal sufficient time (approximately 2 minutes) to settle before recording the data. We recorded the coincidence counts data for a total of 200 iterations with an integration time of only 50 ms to enhance the accuracy of the measurement. Using these coincidence counts and comparing them with the HOM curve (see Fig. \ref{Figure 2}(b)), we estimate the optical delay as a function of the crystal temperature. The results are shown in Fig. \ref{Figure 3}.

As evident from Fig. \ref{Figure 3}(a), the optical delay due to the change in the group index of the sample (PPKTP crystal) increases linearly with the crystal temperature in the range of 26 - 29$^{\circ}$C with a slope of 1.158 $\upmu$m$^{\circ}$C$^{-1}$ of the crystal temperature. Since the current HOM interferometer can measure a static delay of 60 nm\cite{singh2023AQT}, using the above slope, we can in-principle measure the group index of a 30 mm long PPKTP crystal for a temperature change as low as 0.05$^{\circ}$C. The increase in the temperature range requires a further increase in crystal temperature, pushing the optical delay out of the linearity range and providing unreliable results. However, by adjusting the initial point, we can measure the group index of the nonlinear crystal over a range of 3$^{\circ}$C. To confirm the reliability of the experiment, we recorded the variation of coincidence counts over 10 minutes for each step change (0.1$^{\circ}$C) of the crystal temperature across 27.1 - 28.0$^{\circ}$C. The results are shown in Fig. \ref{Figure 3}(b). It is evident from Fig. \ref{Figure 3}(b) that a 0.1$^\circ$C change in crystal temperature results in a substantial change in coincidence counts, exceeding 100 counts per 100 ms of integration time. This considerable change, well above the dark count and accidental counts, establishes the feasibility of resolving group index induced optical delays for a temperature change smaller than 0.1$^\circ$C in real-time operation. We could not access the lower step of the temperature change due to the unavailability of a suitable oven in our lab. As observed previously \cite{singh2023AQT}, the temporal variation of the coincidence counts can be attributed to various external perturbations, including the laser intensity fluctuation, air current, and local temperature instability in the laboratory. However, it is intriguing to observe that the change in coincidence counts resulting from various external perturbations is considerably smaller than the change in coincidence counts due to the alteration in the group index of the sample (PPKTP crystal).
\begin{figure}[ht]
    \centering
    \includegraphics[width=\linewidth]{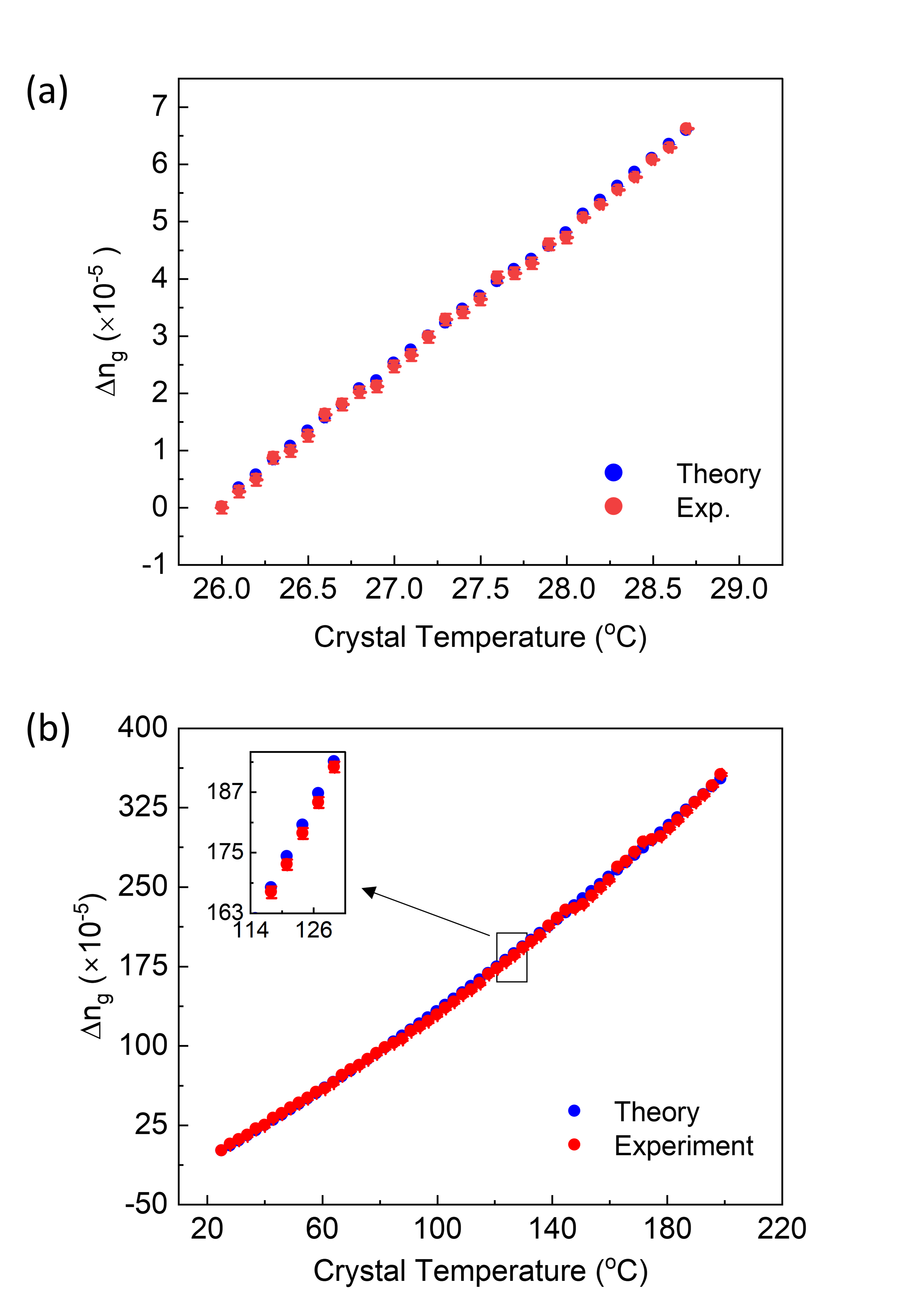}
    \caption{Variation of group index of the PPKTP crystal (S) with temperature measured using a) optical delay-dependent coincidence variation in the linear region of the HOM curve, and b) optical delay mirror, compensating the coincidence change mediated by the temperature-dependent group index variation. (Inset) shows the magnified image of the temperature-dependent group index variation. Blue and red dots are theoretical and experimental results, respectively. }
    \label{Figure 4}
\end{figure}

Using the optical delay corresponding to the change in sample crystal (L = 30.12 mm) temperature from Fig. \ref{Figure 3}(a), the group index value from Table \ref{table1} and the thermal expansion of the sample crystal, $\Delta L(\Delta T)$ from \cite{emanueli2003} in Eq. \ref{eq6}, we calculated the temperature-dependent group index, $\Delta n_{g}$($\Delta$ T), of the PPKTP crystal (S). The results are shown in Fig. \ref{Figure 4}. The integration time for these measurements is set to 50 ms. As evident from Fig. \ref{Figure 4}(a), the temperature-dependent group index (red dots) measured using the HOM curve of FWHM width around 4.11 $\upmu$m has a temperature range of 3$^{\circ}$C at any initial temperature and the corresponding measurement range for group index variation of $\sim$7.3 $\times$ 10$^{-5}$. The measurement of group index for the minimum change of crystal temperature, $0.1^{\circ}$C, sets the resolution of the HOM-based quantum sensor to be $\sim 2.25 \times 10^{-6}$ for sample length of 30.12 mm. Using the dispersion relations of the PPKTP crystal from Ref. \cite{kato2002, emanueli2003}, we calculated the temperature-dependent group index variation (blue dots) to be in close agreement with the experimental results (red dots). As compared to previous report on high precision group index measurements \cite{reisner2022} of a sample (optical fiber) length of 50 cm having precision of 3 $\times 10^{-5}$ cm$^{-1}$ (absolute value of 6 $\times 10^{-7}$ for 50 cm sample) our measurement precision, 6.75 $\times 10^{-6}$ cm$^{-1}$, show an unprecedented enhancement by $\sim$ 444\%. Such enhancement was possible due to the balance between the high photon pair rate and the narrow HOM-curve resulting from the control of the length of the nonlinear crystal producing broadband photon pairs \cite{singh2023AQT}. 

To evaluate group index variations across an extended temperature range, one can utilize a HOM curve with a higher full width at half-maximum (FWHM) width. However, as initially explained, this approach typically leads to poor precision in group index measurement. To circumvent range-dependent resolution issues and enhance the measurement range while maintaining high resolution ($\sim 2.25 \times 10^{-6}$), we adjusted the delay stage to the start point (see Fig. \ref{Figure 2}(b) for the initial crystal temperature of 26$^{\circ}$C and recorded the coincidence counts for any integration time of 100 ms. Knowing the coincidence counts for this initial point, we adjusted the optical delay stage and compensated for the change in the coincidence count arising from the temperature-dependent group index change. Since the entire HOM curve is not utilized for this experiment, we can access an unlimited range restricted by the range of the delay stage with the finest accuracy determined by the narrow HOM curve. We increased the crystal temperature with a step of 3 $^{\circ}$C and compensated the corresponding optical delay by moving the delay stage, thus accessing an unlimited working range with high accuracy.
In the current experiment, we measured the group index variation of the PPKTP crystal from room temperature to the allowed maximum temperature of 200$^\circ$C with a precision of $\sim 2.25 \times 10^{-6}$. The experimental results are shown in Fig. \ref{Figure 4}(b). As evident from Fig. \ref{Figure 4}(b), the group index of the PPKTP crystal changes by $\sim 3.5 \times 10^{-3}$ for the increase of temperature to 200$^\circ$C. The magnified image, as shown by the inset of Fig. \ref{Figure 4}(b), shows the resolution of the measurement to be in the range of $\sim 2.25 \times 10^{-6}$. The close agreement between the calculated values (blue dots) and experimental results (red dots) signifies the reliability of this delay compensation technique for conducting group index measurements over an extended range with notable precision of $\sim 2.25 \times 10^{-6}$. The effective optical path length change for this measurement can be calculated from the total temperature change ($\Delta$T = 175$^\circ$C) and the slope of optical delay versus crystal temperature (1.158 $\upmu$m/$^\circ$C), resulting in an approximate change of $\sim$200 µm. We could not increase the temperature beyond to avoid thermal damage to the crystal.\\

\section{Conclusion}
In conclusion, we have successfully demonstrated a HOM interferometer-based quantum sensor for measuring group index variations of nonlinear crystals with high precision. Using the HOM curve of FWHM width of 4.11 $\pm$ 0.01 $\upmu$m resulting from the photon pairs generated by a PPKTP crystal of length 1 mm, we have measured the temperature-dependent group index of PPKTP crystal over a temperature range of 3$^{\circ}$C at any arbitrary initial temperature with a resolution of 6.75 $\times 10^{-6}$ per centimeter length of the sample. This resolution is $>$400$\%$ better than the previous group index measurement \cite{reisner2022}. Such enhancement was possible due to the use of the optimum length of the nonlinear crystal to control the spectral bandwidth of the {photon-pair}, thus, the width of the HOM curve, while maintaining a high generation rate of photon pairs to ensure fast measurement. Using compensation of group index mediated optical delay by the linear optical delay stage, we observed the possibility of measuring group index over a measurement range of 175$^{\circ}$C with high precision of $\sim 2.25 \times 10^{-6}$. The current demonstration opens the possibility of HOM-based quantum sensors for quantum optical coherence tomography with high resolution and measurement range.

\section*{ACKNOWLEDGMENTS}
All authors acknowledge the support of the Dept. of Space, Govt. of India. G. K. S. acknowledges the support of Department of Science and Technology, Govt. of India through Technology Development Program (Project DST/TDT/TDP-03/2022).

\section*{AUTHOR DECLARATIONS}
\subsection*{Conflict of Interest}
The authors have no conflicts to disclose.
\subsection*{Data Availability Statement}
The data that support the findings of this study are available from the corresponding author upon reasonable request.

\section*{References}
\bibliography{aipsamp}

\end{document}